\documentclass[12pt]{article} 
\usepackage{graphicx} 
 
\newsavebox{\sboxpubnumber} 
\newsavebox{\sboxpubdate} 
\newcommand{\pubdate}[1]{\begin{lrbox}{\sboxpubdate}{#1}\end{lrbox}} 
\newcommand{\pubnumber}[1]{\begin{lrbox}{\sboxpubnumber}{\begin{tabular}{l} #1 \\ 
				 \usebox{\sboxpubdate} 
				 \end{tabular}} 
                           \end{lrbox} 
                           \pubblock} 
\newcommand{\Title}[1]{\begin{center} {\Large #1 } \end{center}} 
\newcommand{\Author}[1]{\begin{center}{ \sc #1} \end{center}} 
\newcommand{\Address}[1]{\begin{center}{ \it #1} \end{center}} 
\newcommand{\andauth}{\begin{center}{and} \end{center}} 
\newcommand{\pubblock}{\rightline{ 
			\usebox{\sboxpubnumber}}} 
\newenvironment{Abstract}{\begin{quotation}  }{\end{quotation}} 
\newenvironment{Presented}{\begin{quotation} \begin{center} 
             PRESENTED AT\end{center}\bigskip 
      \begin{center}\begin{large}}{\end{large}\end{center} 
      \end{quotation}} 
 
\begin{document} 
 
\begin{titlepage} 
\pubdate{\today}                    
\pubnumber{XXX-XXXXX \\ YYY-YYYYYY} 
 
\vfill 
\Title{Constraining Cosmic Quintessence} 
\vfill 
\Author{G. J. Mathews}
\Address{Center for Astrophysics, Department of Physics, University of Notre Dame\\ 
         Notre Dame, IN 46556  USA} 
\Author{
 K. Ichiki$^{1,2}$, T. Kajino$^{1,2,3}$ and  M. Orito$^1$
}
\Address{
$^1$National Astronomical Observatory, 2-21-1, Osawa, Mitaka, Tokyo 181-8588, 
Japan}
\Address{
$^2$University of Tokyo, Department of Astronomy, 7-3-1
Hongo, Bunkyo-ku, Tokyo 113-0033, Japan }
\Address{
$^3$Graduate University for Advanced Studies, Dept.
of Astronomical Science, 2-21-1, Osawa, Mitaka, Tokyo 181-8588, Japan}
\andauth 
\Author{M. Yahiro }
\Address{
Department of Physics and Earth Sciences, University of the Ryukyus,
Nishihara-chou, Okinawa 903-0213, Japan }
\vfill 
\begin{Abstract} 
In quintessence models a scalar field couples with the dominant
constituent and only acts like a cosmological constant 
after the onset of the matter dominated epoch.  
A generic feature of such solutions, however, is the possibility of
significant energy density in the scalar field during the
radiation dominated epoch.  This possibility is even greater if the
quintessence  field begins in a kinetic-dominated  regime, for example
as might be generated at the end of "quintessential
inflation."  As such, these models can
be constrained by primordial nucleosynthesis
and the epoch of photon decoupling.
Here, we analyze both kinetic dominated and power-law 
quintessence fields (with and without a supergravity correction).
We quantify the allowed initial conditions and effective-potential parameters.
We also deduce constraints on the epoch of matter creation at the end
of quintessential inflation.
\end{Abstract} 
\vfill 
\begin{Presented} 
    COSMO-01 \\ 
    Rovaniemi, Finland, \\ 
    August 29 -- September 4, 2001 
\end{Presented} 
\vfill 
\end{titlepage} 
\def\thefootnote{\fnsymbol{footnote}} 
\setcounter{footnote}{0} 
 
\section{Introduction} 

Observations \cite{garnavich,perlmutter} of Type Ia supernovae  and 
the power spectrum of the cosmic microwave background,
 together with complementary observational constraints \cite{wang},
all indicate  
 that the universe is presently accelerating.  In one interpretation,
this acceleration is due to
the influence of a dominant dark energy with a negative pressure.
The simplest candidate  
for such dark energy is a cosmological constant 
for which the equation of
state is $\omega \equiv P/\rho  = -1$.  
A second possibility is derived from the so-called ``quintessence" models,
in which the dark   energy is the result of a scalar field 
$Q$ slowly evolving along an effective
potential $V(Q)$.  The equation of state 
is negative $-1 \le w_Q \le 0$,
but not necessarily constant. 

Introducing a quintessence field helps to
reconcile the
fine tuning problem 
and the cosmic coincidence problem 
associated with a simple cosmological constant.
\cite{wetterich}-\cite{yahiro}.
Specific forms of the quintessence effective 
potential can be chosen such that 
the field $Q$ evolves
according to an attractor-like solution.
Thus, for a wide variety of initial conditions,
the solutions for $Q$ and $\dot Q$  rapidly approach a common
evolutionary track. 
These solutions
lead naturally to a cross over from an earlier 
radiation-dominated solution to one in which the universe 
is dominated by dark  energy at late times.
Another interesting possible feature is that such models might
naturally arise during matter creation at the
end of an  earlier  ''quintessential`` inflationary epoch 
\cite{vilenkin}.
In this case,  the $Q$ field emerges in a 
kinetic-dominated regime at energy densities well above the tracker 
solution.

It is not yet clear, however, that these models
have altogether solved the fine-tuning and cosmic-coincidence problems
\cite{kolda99,copeland}, 
and there may be some
difficulty in embedding quintessence models in string theory \cite{hellerman}.
Nevertheless, several such tracker fields 
have been proposed \cite{brax}
whose effective potentials
may be suggested by particle physics models
with dynamical symmetry breaking, by nonperturbative effects
\cite{zlatev}, by generic kinetic terms "$k$-essence"  in an  effective
action describing the moduli and massless degrees of freedom in string and
supergravity theories \cite{chiba}-\cite{armendariz2},  
or by static and moving branes in
a dilaton gravity background \cite{chenlin}.  
 
A general feature of all such solutions,
however, is the possibility for a significant
contribution of the $Q$ field to the total energy density during the
radiation-dominated  epoch as well as
the present matter-dominated epoch.  The
yields of primordial nucleosynthesis  and the power spectrum
of the CMB can be strongly affected by
this background energy density.
Therefore, we utilize primordial nucleosynthesis
and the CMB power spectrum to constrain viable quintessence models.

\section{Quintessence Field}
 A variety of quintessence  effective potentials \cite{brax} 
or $k$-essence effective actions \cite{chiba}-\cite{armendariz2}
can be found in the literature.  
Observational constraints on such quintessence models have been
of considerable recent interest \cite{yahiro,bean,chen}.
  In this paper, we  
 describe the work presented in \cite{yahiro} on
$Q$-field and/or kinetic-dominated quintessence models.
We concentrate  on
an inverse power law for the $Q$ field as
originally proposed by Ratra and Peebles \cite{ratra},
$V(Q) = M^{(4  +  \alpha)} Q^{-\alpha}$, 
where, $M$  and $\alpha$ are parameters.
The parameter $M$ in these potentials is fixed by the
condition that $\Omega_Q = 0.7$ at present.
Therefore, 
$\rho_Q(0) = 0.7 \rho_c(0) = 5.7 h^2 \times 10^{-47} {~\rm GeV}^4$,
 and
$M  \approx  (\rho_Q(0)Q^\alpha)^{1/(\alpha +4)}$.
If $Q$ is presently near the Planck mass and $\alpha$ is not too small
(say $\alpha ^>_\sim 4$), this 
implies  a reasonable value \cite{zlatev} for $M$  which 
resolves the fine tuning problem.  

 We also consider a modified  form of $V(Q)$ as proposed
by \cite{sugraref} based upon the condition that
the quintessence fields be part of supergravity models.
The rewriting of the effective potential in supergravity
depends upon choices of the K\"ahler potential \cite{copeland}.
The flat K\"ahler potential yields an extra factor of
$exp{\{3Q^2/2m_{pl}^2\}}$ \cite{sugraref}.  This comes about
by imposing the condition that the expectation 
value of the superpotential vanishes.
The Ratra potential thus becomes 
$V_{SUGRA}(Q) \rightarrow  
M^{(4 + \alpha)} Q^{-\alpha} \times exp{(3Q^2/2m_{pl}^2)}$,
where the exponential correction becomes largest near the present time
as $Q \rightarrow  m_{pl}$.  
This  supergravity motivated  effective potential is known as
the  SUGRA potential.  The fact that this potential 
has a minimum for $Q = \sqrt{\alpha/3} m_{pl}$ changes the dynamics.
It causes the present value of $w_Q$ to evolve to a
cosmological constant ($w_Q \approx -1$) much quicker than for the bare
power-law potential \cite{brax}.

The quintessence field $Q$ obeys the equation of motion
$\ddot Q  + 3 H \dot Q  + d V/dQ  = 0$,
where the Hubble parameter $H$ is given from the solution to the
Friedmann equation,
$H^2  = ({\dot a / a})^2  =  {1 / m_{pl}^2} 
(\rho_B +  \rho_Q)$,
where $m_{pl} = (8\pi G/3)^{-1/2} = 4.2 \times 10^{18}$ GeV,
$\rho_B$ is the energy density in background radiation and matter,
and  $a$ is the cosmic scale factor.
As the $Q$ field evolves, its contribution to the energy density
is given by
$\rho_Q =  {\dot Q^2 / 2}   +  V(Q)$. Similarly, 
the pressure in the $Q$ field is
$P_Q =  \dot Q^2/2 - V(Q)$.
 The equation of state parameter  $w_Q$ is a time-dependent
quantity,  $w_Q  \equiv P_Q/\rho_Q = 1 - 2 V(Q)/\rho_Q$,
where the time dependence derives from the evolution of $V(Q)$ and
$\rho(Q)$.

\section{Nucleosynthesis Constraint}
The quintessence initial conditions are  probably set in
place near the inflation epoch.  By the time of the 
big bang nucleosynthesis (BBN) epoch,
many of the possible initial conditions will have
already achieved the tracker solution.
However, for initial conditions sufficiently removed
from the tracker solution, it is possible that 
the tracker solution has not yet been  achieved by the time of BBN at
$ 0.01 ^<_\sim T ^<_\sim 1$ MeV,
$10^{8}~^<_\sim z~^<_\sim 10^{10}$.

For many possible initial conditions 
the tracker solution is obtained before nucleosynthesis.
Along the tracker solution, $\rho_Q$
diminishes in a slightly different 
way than the radiation-dominant background energy density. 
For example, as long as $\rho_Q << \rho_B$,
the $Q$-field  decays as
 $\rho_Q \propto  a^{-3(1+w_Q)}$, 
with  
$w_Q = (\alpha w_B - 2)/(\alpha+2) < w_B$. 
The equation of state $w_Q$ is  only equal
to the background equation of state
$w_B$ in the limit $\alpha \to \infty$.  
Nevertheless, the tracker solution does not deviate much
from $\rho_B$, even at high redshift for most values  of
$\alpha$.
Hence, one can characterize the nucleosynthesis results
by the (nearly constant)  ratio  $\rho_Q/\rho_B$
during the BBN epoch.
If the energy density in the tracker solution
is close to the background energy density,
the nucleosynthesis will be affected by
the increased expansion rate
from the increased total energy density. Such a situation occurs for
large values of the power-law exponent $\alpha$.

A second possibility is
that the energy density  $\rho_Q$ could exceed the tracker solution
and be comparable to or greater than the background energy density
during primordial nucleosynthesis.  
The kinetic energy in the $Q$ field then dominates over
the potential energy contribution to $\rho_Q$ and  
$w_Q = +1$ so that the kinetic energy density  
diminishes as $a^{-6}$.
In this case
 there could be a significant contribution from $\rho_Q$
during nucleosynthesis as the $Q$ field approaches the tracker solution.  
The strongest constraints on this case would arise when
$\rho_Q$ is comparable to the background energy density  near the time of
the weak-reaction freese out, while the later nuclear-reaction epoch 
might be unaffected.  
This case is particularly interesting as this kinetic-dominated evolution 
could be generated by an earlier quintessential inflation epoch
\cite{vilenkin} as described below. 

A final possibility might occur if the $Q$ field approaches the
tracker solution from below.
In this case, the tracker solution may be achieved   
after the the BBN epoch so that a small $\rho_Q$ during BBN
is easily guaranteed.  However, the ultimate
tracker curve  might still have
a large energy density at the later CMB epoch as described below.

Adding energy density from the $Q$ field tends to increase
the universal expansion rate.  Consequently, the weak reaction rates
 freeze out at a higher temperature $T_w$.  
This fixes the neutron to proton ratio ($n/p \approx \exp[{(m_p -m_n)/T_w]}$)
at a larger value.  Since most of the free neutrons are converted into
He$^4$, the primordial helium production  is increased.
Also, since the epoch of nuclear reactions is shortened, the
efficiency of burning deuterium into helium is diminished
and the deuterium abundance also  increases.
Hence, very little excess 
energy density from the $Q$ field is allowed.
 The primordial light-element abundances
deduced from observations have been reviewed by a number
of recent papers \cite{osw99}-\cite{tytler00}.
There are several outstanding uncertainties.  For our purposes 
\cite{yahiro} we adopt
         $ 0.226 \leq Y_p \leq 0.247$ and 
        $ 2.9 \times 10^{-5} \leq D/H \leq 4.0 \times 10^{-5}$
as the most relevant constraints.

\begin{figure}[htb]
\centering
    \includegraphics[width=3.8in]{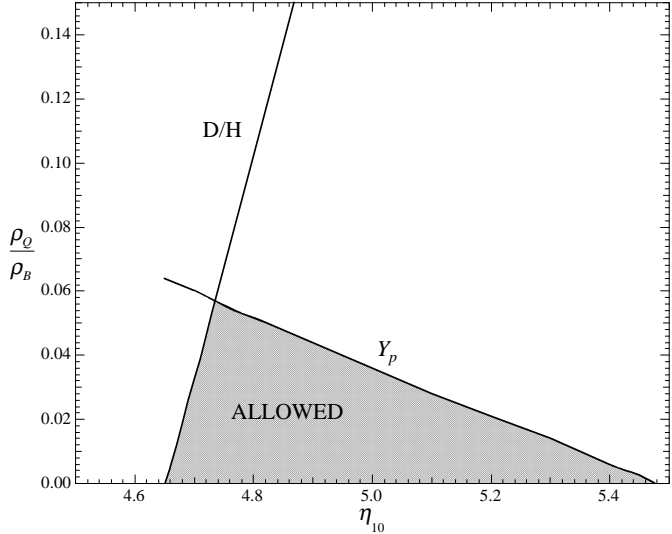}
\caption{Constraints on the ratio of the energy density in the quintessence
field to the background energy density  $\rho_Q/\rho_B$  (at $T = 1$ MeV)
from the primordial abundances as indicated. The allowed region corresponds
to $Y_p \le 0.247$ and D/H $\le 4.0\times 10^{-5}$.}
\protect\label{fig:1}
\end{figure}

Figure \ref{fig:1} summarizes allowed values of $\rho_Q/\rho_B$ at 
$T=$1 MeV  ($z \approx 10^{10}$) 
 based upon the nucleosynthesis constraints.
The region to the right of the curve labeled $D/H$ corresponds
to models in which the primordial deuterium constraint is satisfied,
$D/H \le 4.0 \times 10^{-5}$.  The region below the line labeled
$Y_p$ corresponds to $Y_p \le 0.247$.  The hatched region summarizes
allowed values for the energy density in the quintessence field
during the nucleosynthesis epoch.

In the present work we deduce an absolute
upper limit of 5.6\%  of the background radiation energy density
allowed in the quintessence field.  This maximum contribution
is only allowed  for $\eta_{10}  \approx  4.75$ or $\Omega_b h^2
\approx  0.017$.   
A smaller contribution is allowed for
other values of $\eta_{10}$.  Indeed, this optimum $\eta_{10}$ value is
$4 \sigma$ 
less than the value implied by the cosmic deuterium abundance
\cite{nollett00,tytler00} 
$\Omega_b h^2 = 0.020 \pm 0.001~(1 \sigma)$ ($\eta_{10} = 
5.46 \pm 0.27$). 
The independent determinations of $\Omega_b h^2$
from  high-resolution
measurements of the power spectrum of fluctuations in the 
cosmic microwave
background favor a value even higher.
Both the BOOMERANG  \cite{boom} and DASI  \cite{dasi}  data  sets imply
$\Omega_b h^2 = 0.022^{+ 0.004}_{-0.003} ~(1 \sigma)$ ($\eta_{10} =
6.00^{+ 1.10}_{-0.81}$).   The deuterium and CMB constraints together
demand that $\eta_{10} \ge 5.19$ which would limit the 
allowed contribution from the $Q$ field to $\le 2\%$ of the background energy
density.

The most restrictive CMB constraint  on $\Omega_b h^2$ derives from
demanding a flat universe ($\Omega_{tot} =  1.0$),
and marginalizing the likelihood function
over all parameters with assumed Gaussian
priors \cite{boom} based upon 
measurements of large-scale structure and Type Ia supernovae.  
 This gives $\Omega_b h^2 = 0.023 \pm 0.003~(1 \sigma)$.
If one adopts this  as a most extreme case, then
  $\Omega_b h^2 \ge 0.020$. This
would correspond to  $\eta_{10} \ge 5.46$.
From Figure 1 this would imply a much more stringent constraint 
that only about 0.1\% of the background energy density could 
be contributed by the $Q$ field.
Of course, this is only a $1\sigma$ constraint, and
the upper limit
to $Y_p$ is not  well enough established to rule out a contribution
to the energy density at the 0.1\% level.
We adopt the conservative constraint of 5.6\%.  Nevertheless,
it is of interest to explore how the quintessence parameters
allowed by BBN might improve should
the constraints from BBN ever be so tightly 
defined.   Therefore, we will consider
 0.1\% as a conceivable limit that demonstrates the sensitivity to
BBN.   Even the most conservative  5.6\% limit  adopted here 
corresponds to only about half of the energy density allowed
in \cite{freese} for 3 neutrino flavors.

\section{Equation of State Constraint}

The $Q$ field  must behave like
dark  energy during the present matter-dominated epoch,
i.e.~the equation of state should be sufficiently negative, 
($w_Q \equiv P_Q/\rho_Q < 0$) by the present time.
 We adopt observational constraints on $w_Q$ 
from \cite{wang}. They adopted a 
most conservative approach based upon progressively less reliable data sets.
Using the most reliable current low-redshift and microwave background
measurements, the 2$\sigma$ limits are $-1 < w_Q < -0.2$.
Factoring in the constraint from Type Ia supernovae reduces
the range to $-1 < w_Q < -0.4$.  This
derives from a concordance analysis of
models consistent with each observational
constraint at the $2\sigma$ level or better. 
A combined maximum likelihood analysis suggests
a smaller range of $-0.8 < w_Q < -0.6$  for quintessence models 
which follow the tracker solution,
though $w_Q \approx  -1$ is still allowed in models with nearly a constant
dark   energy. We invoke these three
possible limit ranges for the present value of $w_Q$.

\section{CMB Constraint}

There are two effects on the epoch of photon decoupling to be considered. 
In the case where the energy density in the quintessence field
is negligible during photon decoupling \cite{brax}
the only effect of the dark   energy is to modify the
angular distance-redshift relation \cite{hu}.  The existence of dark   energy
along the look-back to the surface of last scattering shifts the 
acoustic peaks in the CMB power spectrum to smaller angular scales
and larger $l$-values.
The amplitude of the first acoustic peak in the power spectrum also increases,
but not as much for quintessence  models  as for a
true cosmological constant $\Lambda$.  The basic features of the observed
power spectrum \cite{boom} can be fit \cite{brax} with either
of the quintessence potentials considered here.
For our purposes, this look-back  constraint
 is already satisfied by demanding that 
$\Omega_Q = 0.7$ at the present time.

The second effect occurs 
if the energy density in the $Q$ field is a significant fraction of the 
background energy during the epoch of
photon decoupling. Then it can increase
the expansion rate and push the $l$ values
for the acoustic peaks to larger values and increase  
their amplitude \cite{hu}.  Such an effect 
has been considered by a number of authors  in various  
contexts  \cite{bean,hu,hannestad}.
For our purposes, we adopt a constraint 
\cite{bean} based upon the latest CMB sky maps of
the Boomerang \cite{boom} and DASI \cite{dasi} collaborations.    
The density in the $Q$ field can not exceed $\Omega_Q \le 0.39$
during the epoch of photon decoupling.  This implies
a maximum of $\rho_Q/\rho_B = \Omega_Q/(1-\Omega_Q)~^<_\sim 0.64$
during photon decoupling. 

\section{Quintessential Inflation}

Another possible constraint arises  if the kinetic term 
dominates at an early stage.
  In this case $\rho_Q \approx \dot Q^2/2$ 
and $\rho_Q$ decreases with scale factor as $a^{-6}$. 
At very early times this kinetic regime can be produced by  
so-called "quintessential inflation" \cite{vilenkin}.
In this paradigm entropy in the matter fields comes from gravitational
particle production at the end of inflation.  
The universe is presumed to exit from inflation directly 
into a  kinetic-dominated quintessence regime during which 
the matter is generated.
An unavoidable consequence of this process, however,
is the generation of gravitational waves along with matter and the 
quanta of the quintessence field 
\cite{vilenkin,riazuelo,giovannini1,ford}
at the end of inflation.

\subsection{Energy in Quanta and Gravity Waves}
The energy density in created particles can be deduced using
quantum field theory in curved space-time 
\cite{vilenkin,giovannini1,ford},
$\rho_B \approx (1/128)  
N_s H_1^4 ({z+1 / z_1 +1})^4 
(g_1 / g_{eff}(z))^{1/3}$,
where $H_1$ and $z_1$ are the expansion factor and redshift at the 
matter thermalization epoch at the end of quintessential inflation, 
respectively.  The factor of 128 in this expression comes from the
explicit integration of the particle creation terms. 

When the gravitons and quanta of the  $Q$ 
field are formed  at the end of inflation,
one expects \cite{vilenkin}
the energy density in gravity waves  to be twice
the energy density in the $Q$-field quanta (because there are two graviton
polarization states).  In this paradigm then, wherever we have deduced a constraint
on $\rho_Q$, it should be taken as the sum of three different contributions.
One is the dark energy from the vacuum expectation value $\langle \rho_Q \rangle$
 of the $Q$ field; a second is from the
fluctuating part $\rho_{\delta Q}$ of the $Q$ field; and a third is
from the energy density $\rho_{GW}$ in relic gravity waves.  Thus, we have
 $\rho_Q \rightarrow (\langle \rho_Q \rangle + \rho_{\delta Q} + \rho_{GW})$. 
The energy density in gravity waves and quanta scales like radiation after
inflation, $\rho_{GW} + \rho_{\delta Q} \propto a^{-4}$, 
while  the quintessence field vacuum expectation value
evolves as
$\langle \rho_Q \rangle \propto a^{-6}$  during the kinetic dominated epoch. 
This epoch following inflation lasts until the energy in the $Q$ field
falls below the energy in background radiation and matter,
$\rho_Q \le \rho_B$.

Thus, for the kinetic dominated initial conditions, 
 gravity waves could be an important contributor
to the excess energy density during nucleosynthesis.
The relative contribution of gravity waves and quintessence quanta
compared to the background matter fields is just given by the
relative number of degrees of freedom.  At the end of inflation, 
the relative fraction of energy density in quanta and gravity
waves is given by  \cite{vilenkin}
$(\rho_{\delta Q} + \rho_{GW})/\rho_B = 3/N_s$, 
where $N_s$ is the number of ordinary scalar fields at the
end of inflation.  In the minimal supersymmetric
model $N_s = 104$.   
Propagating this ratio to the time of nucleosynthesis requires another
factor of  $(g_n/g_{1})^{1/3}$ where $g_n = 10.75$ counts the number of
effective
relativistic degrees of freedom just before electron-positron annihilation,
and $g_{1}$ counts the number of degrees of freedom 
during matter thermalization after the end of inflation.
In the minimal standard model  $g_{1} = 106.75$, but in supergravity
models this increases to $\sim 10^3$.

 Combining these factors we have
$(\rho_{\delta Q} + \rho_{GW})/\rho_B \le 0.014$,
during nucleosynthesis.  Hence, in this paradigm, the allowed values
of $\rho_Q/\rho_B$ consistent with nucleosynthesis 
could be reduced from a maximum of 0.056 to 0.042,
further tightening the constraints deduced here.

\subsection{Gravity-Wave Spectrum}

There has been considerable recent interest
\cite{riazuelo,giovannini1} 
in the spectrum of gravity waves
produced in the quintessential inflation paradigm.
One might expect that the {\it COBE} constraints on the spectrum 
also lead to constraints on the $Q$ field.  However,
we conclude below that no significant constraint on the
initial $\rho_Q$ or effective potential is
derived from the gravity wave spectrum.  On the other hand,
the BBN and CMB gravity-wave  constraints 
can be used to provide useful constraints on the quintessential
inflation epoch as we now describe.
Our argument is as follows:
The  logarithmic gravity-wave energy  spectrum observed
at the present time can be defined in terms 
of a differential closure fraction,
$\Omega_{GW}(\nu) \equiv (1 /\rho_c) (d \rho_{GW} / d \ln{\nu})$,  
where the  $\rho_{GW}$ is the present energy density in relic gravitons 
and $\rho_c(0) = 3 H_0^2/8 \pi G = H_0^2 m_{pl}^2$ is the critical density. 
This spectrum  has been derived       
in several recent papers \cite{riazuelo,giovannini1}.  
It is characterized
by spikes at low and high frequency.  The most
stringent constraint at present derives from 
the {\it COBE} limit on the tensor component
of the CMB power spectrum  at low multipoles.  There is also a weak constraint from
the binary pulsar \cite{giovannini1} and an integral constraint
 from nucleosynthesis as mentioned above.

 For our purposes, the only possible
new constraint comes from the {\it COBE} limits
on the tensor component of the CMB power spectrum.  
The soft branch in the gravity-wave spectrum lies in the frequency range
between the present horizon 
$\nu_0 = 1.1 \times 10^{-18} \Omega_{M}^{1/2}h $ 
Hz and the decoupling frequency $\nu_{dec}(0) = 1.65 \times 10^{-16} 
\Omega_{M}^{1/2} h $  Hz,
where  we adopt $\Omega_{M} = 0.3$ for the  present matter
closure fraction.
The  constraint on the spectrum  can be
written \cite{giovannini1},
\begin{equation}
\Omega_{GW}(\nu) = \Omega_\gamma  {81 \over (16 \pi)^2} \biggl(
{g_{dec} \over g_{1}}\biggr)^{1/3}
\biggl({H_1 \over m_{pl}}\biggr)^2 
\times \biggl({\nu_{dec} \over \nu}\biggr)^2 
\times 
\ln^2{\biggl({\nu_{r} \over \nu_1}\biggr)} \le  6.9 h^{-2} \times 10^{-11} ~~,
\label{gwspec}
\end{equation}
where, $\Omega_\gamma = 2.6 \times 10^{-5} h^2$ is the present closure 
fraction in photons.  The number of relativistic degrees of freedom at
decoupling is $g_{dec} = 3.36$.  As noted previously,
$g_1$ is the number of relativistic degrees of
freedom after matter thermalization.  In the minimal
standard model is $g_{1} = 106.75$. The quantity $H_1$ is the expansion rate at 
the end of inflation.  It is  simply
related to the kinetic energy $\rho_Q$ after inflation, 
$\rho_{Q}(z_1) = H_1^2 m_{pl}^2$.
The logarithmic term in Eq.~(\ref{gwspec}) involves the ratio of the  
present values of the  frequency $\nu_r(0)$
characteristic of the start of radiation domination ($\rho_B = \rho_Q$
at $z = z_r$), to  
the frequency characteristic of matter thermalization
at the end of inflation $\nu_1(0)$. This ratio is just
${\nu_r(0) / \nu_1(0)} = ({z_r + 1 / z_1 + 1})^2$. 
The identity $\rho_B = \rho_Q$ at $z = z_r$ then gives,
${\nu_r / \nu_1} =  {N_s/ 128} 
({H_1 / m_{pl}})^2 ({g_1 / g_r})^{1/3}$,  
where for the cases of interest $g_r = 10.75$.

Collecting these terms, we can then use Eq.~(\ref{gwspec}) 
to deduce a constraint on the expansion factor
at the end of inflation
$H_1^2 < 1.4  \times 10^{-11} m_{pl}^2$.  
For kinetic dominated models, $\rho_Q(z_1)$ at the end of inflation
is simply related to the
energy density $\rho_Q$ at $z $,
$\rho_{Q}(z)  = \rho_{Q}(z_1) [(z+1)/(z_1+1)]^6$.  
Similarly the background matter energy density scales as
$\rho_{B}(z) = \rho_{B}(z_1) ({g_{1} /
g_{eff}(z)})^{1/3} [(z +1)/(z_1+1)]^4$. 

Considering the present energy density in photons and neutrinos, we can
find a relation between $H_1$ and $z_1$:
$\rho_{\gamma 0} + \rho_{\nu 0} = 1.1 \times 10^{-125} m_{pl}^4 
 = (N_s / 128) H_1^4 
({g_1 / g_{dec}})^{1/3} (z_1 + 1)^{-4}$.  
We then deduce  that $z_1 < 8.4 \times 10^{25}$ and
there is only a lower
limit on $\rho_Q/\rho_B$ given by the constraint on $H_1$.
At our initial epoch $z = 10^{12}$, we find, $\rho_Q(z)/\rho_B(z)
~^>_\sim  5.6 \times 10^{-18}$.  
This limit is not particularly useful because $\rho_Q$
field must exceed $\rho_B$ at $z_1$ in order
for the gravitational particle production paradigm to work.  
 The implication is  then
that all initial conditions in which the kinetic term dominates over
the background energy at $z_1$ 
are allowed in the quintessential inflation scenario.
Hence, we conclude that the 
gravity-wave spectrum does not presently constrain
the initial $\rho_Q$ or $V(Q)$ in the quintessential inflation model.

However, the limits on $\rho_Q/\rho_B \le 560$ derived from the BBN
constraints discussed below can be used to place a lower limit on the
expansion rate at the end of inflation in this model.  This in turn implies
a lower limit on the redshift for the end of 
quintessential inflation.  Thus, we have
$2.2 \times 10^{-21} < {H_1^2 / m_{pl}^2} < 1.45 \times 10^{-11}$,
and $1.0 \times 10^{21} < z_1 < 8.4 \times 10^{25}$
in the minimal supersymmetric model.

This implies that quintessential inflation must 
end at an energy scale somewhere between about 10$^{8}$ and 10$^{13}$ GeV,
well below the Planck scale.   By similar reasoning one can apply
this argument to the gravity-wave spectrum from normal inflation as
given in \cite{giovannini1}.  We deduce an upper limit to the
epoch of matter thermalization of
$H_1 \le 3.1 \times 10^{-10} m_{pl}^2$ which implies
 $z_1 \le 7.3 \times 10^{28} [g(z_1)/3.36]^{1/12}$.
In this case there is no lower limit from BBN as there is no 
$Q$ field present after inflation.

\section{Results and Discussion}

The equations of motion  were evolved
for a variety of initial $Q$ field strengths and power-law parameters $\alpha$.
As initial conditions, the quintessence field was assumed 
to begin with equipartition,
i.e.~$\dot Q^2/2  = V(Q)$, and $w_Q = 0$.  
This seems a natural and not particularly restrictive choice, since
$w_Q$ quickly evolves toward  the kinetic ($w_Q = +1$)
 or the tracker solution  depending
upon whether one begins above or below the tracker curve.

\begin{figure}[htb]
\centering
 \includegraphics[height=5.0in,angle=-90]{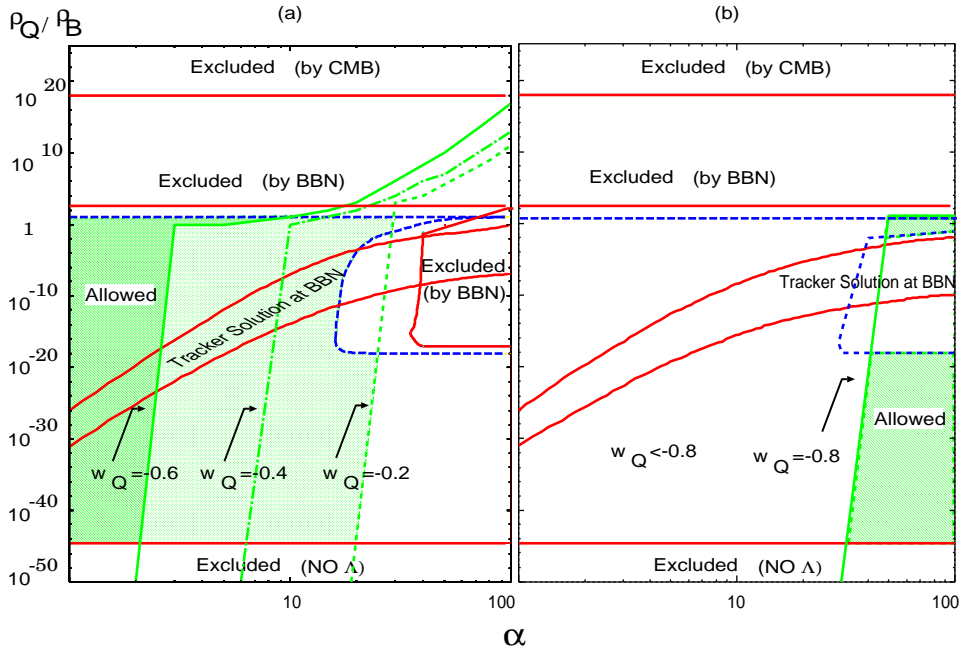}
\caption{
Contours of allowed values for $\alpha$ and initial $\rho_Q/\rho_B$ (at $z = 10^{12}$)
from various constraints as indicated for (a) the bare power-law
potential, and (b) the SUGRA corrected potential.
Models in which the tracker solution is obtained by the BBN epoch are 
indicated by the upper and lower curves. 
Values of $\alpha$ to the right of the lines labeled $w_Q = -0.6,-0.4,-0.2$ 
on (a) are excluded by the requirement that the present equation of
state be sufficiently negative.
The BBN constraint for a maximum energy density in the
$Q$ field of 0.1\% (dotted line)  and 5.6\% (solid line) are also indicated.
 For the SUGRA potential
(b) all tracker solutions to the right of the
region labeled $w_Q = -.8$ are allowed.}
\protect\label{fig:2}
\end{figure}

 Constraints on $\alpha$ and the initial value for the $Q$-field energy density
$\rho_Q(z)/\rho_B(z)$ at $z = 10^{12}$ were deduced numerically.  These
 are summarized in Figure \ref{fig:2} for both: (a) the bare
Ratra power-law  potential;  and  (b) its SUGRA corrected form.
Our constraints can easily be converted to closure fraction
by  $\Omega_Q = (\rho_Q/\rho_B)/ (1 + \rho_Q/\rho_B)$.  
For purposes of illustration, we have arbitrarily 
specified initial conditions at $z = 10^{12}$, corresponding to 
$ T \sim 10^{12}$ K, roughly just after the time of the QCD epoch.
At any time the energy density $\rho_{rel}(z)$ in relativistic particles 
is just
$\rho_{rel}(z) = \rho_{\gamma 0} 
({3.36 / g_{eff}(z)})^{1/3} (z+ 1)^4$,
where $\rho_{\gamma 0} = 2.0 \times 10^{-51} $ GeV$^4$,
is the present energy density in microwave background photons, and
we take  $g_{eff}(z) = 10.75$ between $z = 10^{12}$
and the beginning of BBN just before 
electron-positron annihilation ($z \approx  10^{10}$).

 The envelope of models which obtain a  tracker solution by the
epoch of nucleosynthesis are indicated by upper and lower
curved lines in Figures \ref{fig:2}a and \ref{fig:2}b. 
The general features of these constraints 
are as follows.  If the initial energy density in the $Q$
field is too large,
the tracker solution is not reached by the time of BBN. 
The $Q$-field energy density can
then significantly exceed the background energy  during nucleosynthesis.
This situation
corresponds to the excluded regions on the top of
Figures \ref{fig:2}a and  \ref{fig:2}b.  
All solutions
consistent with the  primordial nucleosynthesis 
 constraints are
also consistent with our adopted CMB $Q$-energy
constraint
 as  also shown at the top of 
Figures  \ref{fig:2}a and \ref{fig:2}b. 

The excluded regions at the top and bottom of figures \ref{fig:2}a 
and \ref{fig:2}b can be easily understood analytically.
For example, the excluded (no $\Lambda$) region at the bottom
of these figures reflects the fact that if $\rho_Q$ is initially below
the value presently required by  $\Omega_\Lambda = 0.7$ it can not
evolve toward a larger value.  Hence, 
$\rho_Q/\rho_B < 2.8 \times 10^{-44}$ is
ruled out for $h = 0.7$.  
Similarly, the ''Excluded by BBN`` region comes from requiring
that $\rho_Q(z_{BBN})/\rho_B(z_{BBN}) < 0.056$.
For this constraint we are only considering cases in which
the $Q$ field is approaching the tracker solution from above 
during nucleosynthesis.  Hence, it is in the kinetic regime in
which $\rho_Q \propto z^6$ while the background scales as $z^4$.  
Thus, we have
 $\rho_Q(z=10^{12})/\rho_B(z=10^{12}) > 0.056 (10^{12}/10^{10})^{6-4}
 = 560$ is excluded.  
By similar reasoning, the ''Excluded by CMB`` region is given
by $\rho_Q(z_{CMB})/\rho_B(z_{CMB}) > 0.64 $,
or $\rho_Q(z=10^{12})/\rho_B(z=10^{12}) >  0.64 (10.75/3.36)^{1/3} 
(10^{12}/10^{3})^{6-4} = 9.4 \times 10^{17}$.

For the bare Ratra  power-law potential (Fig.~2a)  the main constraint is simply
the requirement that the equation of state be sufficiently negative by the 
present time.  
The sensitivity of 
the allowed power-law exponent to the equation of state is indicated
by the $w_Q =$ -0.2, -0.4, and -0.6 lines on Figure \ref{fig:2}a.   
In the present $Q$-dominated
epoch, lines of constant
$w_Q$ must be evaluated numerically.
The slight slope to these curves comes from the fact that 
$V(Q)/\rho(Q)$ has not yet reached unity,
i.e.~there is still some small kinetic contribution to $\rho(Q)$ and
the amount of kinetic contribution depends upon $\alpha$.

For the bare Ratra power-law potential, tracker solutions
with $\alpha ^<_\sim 20$ are allowed if $w_Q \le -0.2$. 
The allowed values for $\alpha$ reduce to $< 9$ and $< 2$
if the more stringent -0.4 and -0.6 constraints are adopted.
However, if $\alpha$ is too small, say $\alpha \le 2$,  then the potential
parameter $M$ becomes a very small fraction of the Planck mass
 and the fine tuning problem is reintroduced.

 For models in which the tracker solution is obtained by the time of
BBN, the potential parameters are only constrained
if the most conservative equation of state limit ($w_Q < -0.2$)
 and most stringent nucleosynthesis constraint 
($\rho_Q/\rho_B < 0.1\%$) are adopted.   On the other hand,
independently of the equation of state constraint,
nucleosynthesis limits a large family of possible kinetic-dominated 
solutions even though they provide the correct present  dark energy.

For the SUGRA-corrected  $Q$ fields (Fig. 2b), 
the constraint from $w_Q$ is greatly relaxed.  In fact,
$w_Q$ is sufficiently
negative ($w_Q < -0.6$) for all $\alpha < 10^4$.
The reason is that all tracker solutions have $w_Q \approx -1$.
This is because $w_Q$ decays much faster toward -1 for
the  SUGRA potential.
Also, the potential has a finite minimum which is 
equal to the present dark-energy density.  
The Q field quickly evolves to near the potential minimum
and has negligible kinetic energy by the present time.
Any potential which becomes flat at late times 
gives $w_Q \approx -1$ and the   dark energy looks 
like a cosmological constant.
All SUGRA models which achieve the tracker solution also have
a small $\rho_Q$ during 
primordial nucleosynthesis.
Hence, there is no constraint from nucleosynthesis except for those
kinetic-dominated models in which the $Q$ field is still far above the tracker
solution during the nucleosynthesis epoch.

We do note, however, that if a lower limit of
$w_Q > -0.8$ is adopted for tracker solutions
 from \cite{steinhardt99},
then only a power law with $\alpha ^>_\sim 30$ is allowed.
  This makes the SUGRA potential the preferred 
candidate for quintessence.
The large $\alpha$ implies values of $M$ 
close to the Planck mass, thus avoiding any fine tuning problem.
However, this potential will be constrainable by BBN
if the light-element constraints become sufficiently precise to 
limit $\rho_Q$ at the 0.1\% level.

\section{Conclusions}

We conclude that for both the bare Rata  inverse power-law potential 
and its  SUGRA-corrected  form, the main constraints for models
which achieve the tracker solution by the nucleosynthesis epoch is
from the requirement that the equation of state parameter becomes
 sufficiently
negative by the present epoch.  The main constraint from nucleosynthesis
is for models which are kinetic dominated ar
the time of nucleosynthesis. 
The SUGRA-corrected potential is the least constrained and
avoids the fine tuning problem for $M$.  Therefore,
it may be the preferred candidate for  the quintessence field,
although BBN may eventually limit this possibility.
We also note that the constraints considered here provide useful
constraints on the regime of matter creation at the end of quintessential 
inflation.


\vskip .1 in
Work supported in part by the Grant-in-Aid for
Scientific Research (10640236, 10044103, 11127220, 12047233)
of the Ministry of Education, Science, Sports, and Culture of Japan,
and also by DoE Nuclear Theory Grant (DE-FG02-95-ER40934 at UND).



\begin{thebibliography}{99} 
 
 
%
%
%
%
%

\bibitem{garnavich} P.~M.  Garnavich, et al.,  Astrophys. J., {\bf 509},
 (1998) 74; Riess, et al., Astron. J, {\bf 116}, (1998) 1009. 
%
\bibitem{perlmutter} S. Perlmutter, et al., Nature {\bf 391}, (1998) 51;
Astrophys. J., {\bf 517}, (1998) 565. 
%
\bibitem{wang}  
L. Wang, R.R. Caldwell, J.P. Ostriker and P. J. Steinhardt, Astrophys. J., 
530, (2000) 17.   
%
\bibitem{wetterich}  
C. Wetterich, Nucl. Phys., B302, (1988) 668.
%
\bibitem{zlatev}  
I. Zlatev, L. Wang, and P. J. Steinhardt, Phys. Rev. Lett., 82, (1999) 896.
%
\bibitem{chiba}  
T. Chiba, T. Okabe, and M. Yamaguchi, Phys. Rev. D62, (2000) 023511.
%
\bibitem{armendariz1}  
C. Armendariz-Picon, V. Mukhanov, and P. J. Steinhardt, Phys. Rev Lett., {\bf 85},
(2000) 4438.
%
\bibitem{armendariz2}  
C. Armendariz-Picon, V. Mukhanov, and P. J. Steinhardt, Phys. Rev D63, 
(2001) 103510.
%
\bibitem{steinhardt99}  
P.  J. Steinhardt, L. Wang and I. Zlatev, Phys. Rev. D59, (1999) 123504.
%
\bibitem{brax}  
P. Brax, J. Martin, and A. Riazuelo,  Phys. Rev D62, (2000) 103505.
%
\bibitem{yahiro}  
M. Yahiro, G. J. Mathews, K. Ichiki, T. Kajino, Phys. Rev. D, (2002) in press.
%
\bibitem{vilenkin}  
P.J.E.~Peebles and A. Vilenkin, Phys. Rev. D59, (1999) 063505.
%
\bibitem{kolda99}  
C. Kolda and D. H. Lyth, Phys. Lett., {\bf B458}, (1999) 197;
S. Weinberg, in Proc. Dark Matter 2000, Marina Del Rey, Feb, 2000, 
astro-ph/0005265, (2000).
%
\bibitem{copeland}  
E. J. Copeland, N. J. Nunes, and F. Rosati, Phys. Rev. {D62}, (2000) 123503.
%
\bibitem{hellerman}  
T. Banks, JHEP submitted (2000), hep-th/0007146; 
T. Banks  and W. Fischler, JHEP submitted (2000), hep-th/0102077; 
S. Hellerman, N. Kaloper, and L. Susskind, JHEP, 06, 003 (2001);
W. Fischler et al., JHEP, 07, 003 (2001) hep-th/0104181;
C. Kolda and W. Lahneman, hep-th/0105300.
%
\bibitem{chenlin} B. Chen and F-L. Lin,  Phys.~Rev. D65, (2002) 044007.
%
\bibitem{bean}  R. Bean, S. H. Hansen, and A. Melchiorri,  Phys.~Rev. D64, (2001) 103508.   
%
\bibitem{chen}  
X. Chen, R. J. Scherrer, and G. Steigman, Phys. Rev D63, (2001) 123504.
%
\bibitem{ratra}  B.~Ratra and P.J.E.~Peebles, Phys.~Rev.~D37, (1988) 3406;
P.J.E.~Peebles and B.~Ratra, Astrophys.~J.~Lett., 325, (1988) L17.
%
\bibitem{sugraref}  
P. Brax and J. Martin, Phys. Lett., {\bf B468}, (1999) 40; Phys.Rev. D61 (2000) 103502;
P. Brax, J. Martin, A. Riazuelo, Phys. Rev. {\bf D62}, (2000) 103505.
%
\bibitem{osw99} K.A.~Olive, G.~Steigman, and T.P.Walker,  Phys. Rep. 333,
(2000) 389.
%
\bibitem{nollett00} K.M.~Nollett and S.~Burles, Phys. Rev. {\bf D61},
(2000) 123505; S.~Burles, K.M.~Nollett, \& M.S.~Turner, Phys. Rev. {\bf D63},
(2001) 063512.
%
\bibitem{tytler00} D.~Tytler, J.~M.~O'Meara, N.~Suzuki, \& D.~Lubin
Phys. Rep., 333, (2000) 409.
%
\bibitem{freese} K.~Freese, F.~Adams, J.~Frieman, and E.~Mottola,
in {\it Origin and Distribution of the Elements}, G.~J.~Mathews, ed.
(World Scientific: Singapore) (1988), pp. 97-115.
%
\bibitem{hu}  
W. Hu, D. Scott, N. Sugiyama, and M. White, Phys. Rev. D52, (1995) 5498.
%
\bibitem{boom} C.B. Netterfield, et al. (Boomerang Collaboration),
Submitted to Astrophys. J. (2001), astro-ph/0104460.
%
\bibitem{dasi} C. Pryke, et al.
(DASI Collaboration), Submitted to Astrophys. J. (2001), astro-ph/0104490.
%
%
\bibitem{hannestad}S.  Hannestad, Phys. Rev. Lett., 85, (2000) 4203.
%
\bibitem{riazuelo}  
A. Riazuelo, and J.-P. Uzan,   Phys. Rev. {\bf D62}, (2000) 083506.
%
\bibitem{giovannini1}  
M. Giovannini, Class. Quant. Grav {\bf 16}, (1999) 2905;
Phys. Rev. D60, (1999) 123511.
%
\bibitem{ford} L. H. Ford, Phys. Rev. {\bf D35}, (1987) 2955.
%
%
%

\end{thebibliography}
\end{document}